\documentclass[aps,prl,nofootinbib,showpacs,twocolumn]{revtex4}%
\usepackage{amsfonts}
\usepackage{amsmath}
\usepackage{amssymb}
\usepackage{graphicx}%

\newcommand{\beq}{\begin{equation}}
\newcommand{\eeq}{\end{equation}}
\newcommand{\beqn}{\begin{eqnarray}}
\newcommand{\eeqn}{\end{eqnarray}}
\newcommand{\bearr}{\begin{array}}
\newcommand{\enarr}{\end{array}}

\begin{document}

%\tightenlines
%\draft

\title{Generic features of the wealth distribution in ideal-gas-like markets}

\author{P. K. Mohanty}
\email[E-mail address: ]{pk.mohanty@saha.ac.in}
\affiliation{Theoretical Condensed Matter Physics Division and Centre for Applied 
Mathematics and Computational Science, Saha Institute of Nuclear Physics, 
1/AF Bidhan Nagar, Kolkata, 700064 India.}

\date{\today}
\vskip 2.cm
\begin{abstract}
We provide an exact solution to the  ideal-gas-like models studied in 
econophysics to understand the microscopic origin of  Pareto-law. 
In these classes of models the key ingredient necessary for  having 
a self-organized scale-free steady-state distribution is the 
trading or collision rule where agents or particles save a definite 
fraction of their wealth or energy and invests the rest for 
trading. Using a Gibbs ensemble approach we could obtain the exact distribution 
of wealth in this model. Moreover we show that in this model (a) good savers 
are always rich and (b) every agent  poor or rich invests the same amount 
for trading. Nonlinear trading rules could alter the generic scenario observed 
here.
\end{abstract}

\maketitle
 
Wealth and its distribution plays an important role in  society. 
Economic policies, natural resources, and human psychology  are certainly important 
factors which govern the  distribution of wealth. However, some features of the distribution are 
independent of the details.  As pointed out by Pareto \cite{Pareto},
a large fraction of the wealth in any  economy is always owned by a 
small fraction of  the population and vice versa. His empirical  formulation,  later   
named as Pareto's law, describes that  the distribution of wealth $w$ follows a power 
law $P(w) = w^{-\gamma}$. Recent studies \cite{Real}  of wealth distribution in several 
countries also confirm that it is indeed the behavior for the "rich", 
which is only about $3\%$ of the population. The rest follow a exponential 
or Gibbs distribution. An interesting analogy \cite{Yako00,Bikas03} has been drawn between the 
economic system and  a system of ideal gas where  particles and their energies are
modeled  as agents and their wealth. Collision between particles is similar to 
trading between agents where energy or wealth is neither created nor destroyed; 
it is just redistributed between agents.  
As pointed out by Yakovenko\cite{Yako00}, such a 
process obviously generates Gibbslike distributions  observed for the majority of 
the population.

  Then what is  the origin of  power-law for the rich?  Chakrabarti and co-workers 
\cite{Bikas03, Mann04}  pointed out that {\it saving } is an important factor   which 
decides and dictates the  distribution for the rich. In a generic society agents have 
different opinions and concepts of saving and, accordingly, each agent 
saves a fraction  of his wealth and invests the rest for trading. 
The available wealth is then shared randomly between two interacting agents. 
These models generically predict  a power-law distribution of wealth with $\gamma=2$.
Later studies indicate that  $\gamma$ is not truly universal and can be changed in certain 
specific cases. A strikingly different distribution of wealth is observed in a system of 
like-minded agents (when saving propensity is the same for every agent), 
where it is  asymmetric and peaked below the average. Numerically it could be well 
fitted to a Gamma distribution \cite{Patri04}; 
however, recent studies indicate  a discrepancy \cite{Rich04}. 
The exact form of the distribution is still an open question and in this paper 
we refer to it as  $Gamma$-like distribution.

It is important to make a distinction between wealth and money, although 
in this paper we have used one as a synonym of the other. 
Wealth is usually refered to things that have economic utility,
$e.g.$, money and property. During an exchange of money for tangible property, 
the wealth of the involved agents does not change. Again, the effective value of 
tangible property in terms of monetary units changes in time violating 
the conservation of total wealth. Such realistic features of economy are not 
included in \cite{Yako00} and \cite{Bikas03, Mann04}; however, these simple minimal models  
remarkably explain  the universal two-class feature of the wealth distributions.
In particular, numerical studies by Chatterjee, Chakrabarti, and Manna (CCM) \cite{Mann04}
clearly suggest an algebraic distribution of wealth   for the rich. 
Later studies (mainly numerical \cite{Book} and a few analytical \cite{Das03}),  
also reveal that algebraic distributions are generic for the 
CCM model and its variants. Exact results, however, are far  from reach. 
In this paper we  will focus on  the exact solution of a generic ideal-gas-like 
economy where saving propensities of agents  are random  and distributed  
arbitrarily. The CCM  model is a special case where the distribution of
saving propensity is uniform.  
Our results clarify why $\gamma=2$ for most cases and also show a way of getting  a 
distributions when $\gamma \ne 2$. We have also pointed out that in these class 
of models  (a) good savers are always rich and (b) every agent invests on the average a fixed amount 
for trading.
 
To be more precise about the model let us 
consider a system of $L$ agents having their saving propensity $0\le\lambda<1$ 
distributed as $g(\lambda)$. The agents are labeled as $i=0,1,\dots,L-1$  such that 
$\lambda_0\le \lambda_1\le\dots\le\lambda_{L-1}$. Let us assume that initially 
total wealth  is randomly  distributed among agents and the average is 
$$\bar E =\frac{1}{N} \sum_i E_i .$$

 A pair of agents, chosen randomly in this model,   exchange their wealth in  the 
following way. Each agent $i$ saves  
$\lambda_i$ fraction of its wealth and  invests  $(1-\lambda_i)$ fraction for 
exchange. The available wealth for the  pair $i$ and $j$ is then shared among the agents in a 
random fashion. Thus,
\begin{eqnarray}
E_i &\to& \lambda_i E_i +  \epsilon_{ij} [ (1-\lambda_i)E_i + (1-\lambda_j)E_j],\cr
E_j &\to &\lambda_j E_j +  (1-\epsilon_{ij}) [ (1-\lambda_i)E_i + (1-\lambda_j)E_j],
\label{eq:E_i}
\end{eqnarray}
where $0<\epsilon_{ij}<1$ is chosen randomly for each trading from a  distribution 
 $h(\epsilon)$  where the average is $r= \langle \epsilon_{ij} \rangle$.

   It is quite evident that when $\lambda_i=0$, the model is identical 
to the  ideal gas model, where particles encounter random elastic collisions.
In this case, irrespective of the initial distribution of energy or wealth,
in equilibrium (after a sufficiently large number of collisions)  wealth 
is redistributed according to Gibbs distribution with a temperature suitably 
defined by $\bar E$. It explains the wealth distribution for a majority which follows 
the exponential law. To understand the origin of  the power-law for the rich,
as explained by the authors of \cite{Bikas03}  we must introduce 
savings. However, note that the particles are no more {\it identical} once the saving 
propensity is different for different agents. Then, one's saving propensity is his 
identity. It is thus appropriate to study the ensemble of such systems. The idea 
of an ensemble is not new in statistical physics. It is a collection of infinitely many mental 
copies of  systems prepared under the same macroscopic conditions. Physical observables 
are then needed to be averaged over the ensemble to get rid of the microscopic 
fluctuations. Let us study an ensemble of $N$ systems labeled by $\alpha=1,\dots , N$, 
prepared with the same average energy $\bar E$.  Thus, in each system 
$\{\lambda_i\}$ are the same whereas the initial wealth is different in 
different systems. Also different is the sequence of pairs $(i,j)$ selected for 
trading and the sharing of available wealth between  the pair ($\epsilon_{ij}^\alpha$) 
during each trading. Thus, it is 
appropriate to find out the distribution of  ensemble-averaged wealth defined by 
\begin{equation}
w_i= \frac{1}{N} \sum_\alpha^N E_i^\alpha,
\end{equation}
where $E_i^\alpha$ is the wealth of an individual $i$ in system $\alpha$. 
In a large system ($L\to \infty$) agent $x=i/L$ has wealth $w(x)$  and 
saving propensity $\lambda(x)$. Since $x$ is uniform in $(0,1)$, 
$\lambda(x)$ can be obtained from the conservation of probability element 
\begin{equation}
g(\lambda) d\lambda = dx.
\label{eq:lambdax}
\end{equation}

    Now we can proceed to derive an effective model  of exchange for $w(x)$.  
First, note that since the pair of agents  are chosen randomly 
for exchange, a specific agent $x$ interacts with different agents in different 
systems in the ensemble. Eventually  agent $x$ interacts with every other agent when 
$N\to \infty$. Second, that for a given pair $x$ and $y$, $\epsilon_{xy}$ is 
different in different systems and  the effective sharing is
$\langle \epsilon_{xy}\rangle= r$. Thus, during trading with $y$, 
the average wealth $w(x)$ of agent $x$ changes to $w^\prime (x;y)$, where
\begin{equation}
w^\prime (x;y)=[ r+\lambda(x)(1-r)] w(x)
+  r[1-\lambda(y)] w(y).
\label{eq1}
\end{equation}
%A similar equation can be writen for changes in $w(y)$.
Since $w(x)$ is stationary in  the steady state, we must have 
\begin{equation}
w(x)= \int_0^1 w^\prime (x;y) dy.
\label{eq2}
\end{equation}
 
 Equation (\ref{eq2}),  together with (\ref{eq1}),  can be solved as 
\begin{equation}
w(x) = \frac{C}{1-\lambda(x)},
\label{eq:wx}
\end{equation}

where $C$ is an arbitrary constant, can be fixed by the average wealth  $\bar E$.
Explicitly, 
%% by the conservation of wealth condition
\begin{equation}
\int_0^{1}  w(x)dx= \bar E.
\label{eq:norm}
\end{equation}
Note that  neither $r$ nor $h(\epsilon)$ appears in Eq. (\ref{eq:wx}).

The distribution of $w$ must satisfy $P(w) dw = dx$; thus
\begin{equation}
P(w) = \frac{dx}{dw} =  \frac{Cg(1-C/w)}{w^2}.
\label{eq:Pw}
\end{equation}
Here the  quantity after the last equality is derived using Eqs. (\ref{eq:wx}) 
and (\ref{eq:lambdax}). It is clear from (\ref{eq:Pw}) that  the asymptotic 
wealth distribution for a generic $g(\lambda)$  is $P(w) \sim w^{-\gamma} $ with $\gamma=2$. 
However one can choose $g(\lambda)=\tilde g (1-\lambda)$ to get a different power law  
$\gamma\ne 2$. For example, when $  g(\lambda) = A (1-\lambda) ^\alpha $ defined 
in the interval $0<\lambda<1$  one gets
\begin{equation}
P(w)=A \frac{C^{1+\alpha}} {w^{2+\alpha}},
\end{equation}
   which has been reported earlier \cite{Bikas03}.  

% Note that the minimum wealth of the system is 
% $C$ and  $P(w)$ is nonzero  for $w>C$. Later we will see that for CCM model $C$ vanishes as 
% $1/\ln(L)$.
  
 To compare  the exact results with the numerical simulations, we must understand
certain existing numerical difficulties of the model. First, that the saving propensity 
is never unity. An agent having saving propensity $\lambda=1$ is a troublesome member 
of the system who never invests and gains  by interacting with other members and 
ultimately owns the whole wealth. In usual numerical simulations,
the maximum saving propensity for  a chosen set $\{\lambda_i\}$ is $q\approx 1$.
For example in CCM model $1-q$ is $O(1/L)$. We must account for this correction 
while evaluating $C$. To do so let us look for a generic system where the saving 
propensity is uniformly distributed in $(0,q)$.  In this case $\lambda(x) =qx$
and using the procedure described here  one gets
 
\begin{equation}
w(x)= \frac{C}{q^{-1} -x} \textrm{ with } ~~~~ C = -\frac{\bar E}{\ln(1-q)}.
\label{eq:wq}
\end{equation}

So, for the CCM model $C \sim \bar E/\ln(L)$. 
Alternately one can calculate $C$ as follows. The CCM model is equivalent to a 
system of $L$ agents with $\lambda_i = i/L$.  In this case, strictly $q=1-1/L$  and 
thus $C= \bar E/\ln(L)$.

\begin{figure}[tbp] 
 \includegraphics*[width=7cm]{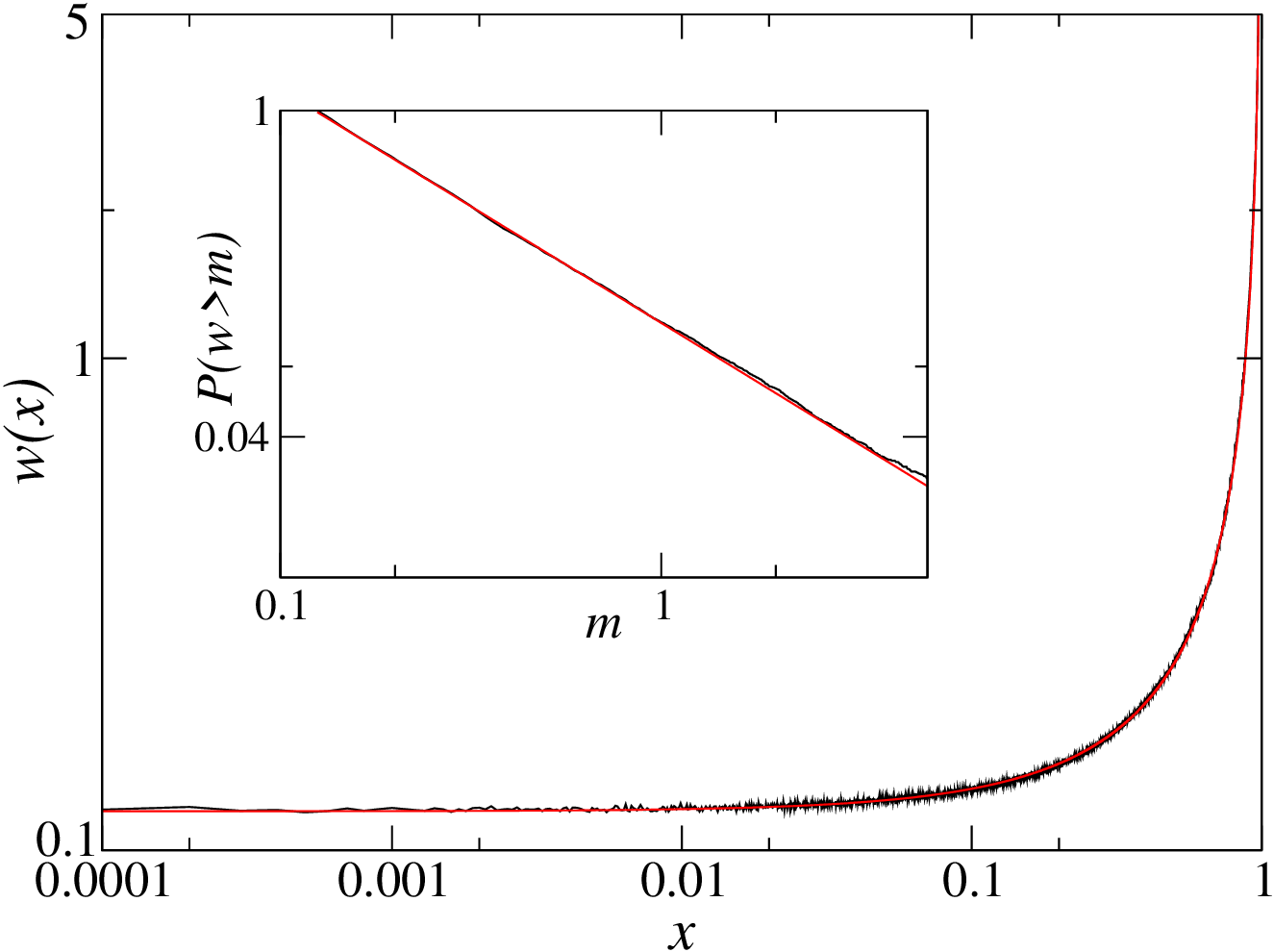}
\caption{Exact results for the CCM model \cite{Mann04} is compared with 
numerical simulations. Practically they are indistinguishable. 
Main figure shows individual wealth of the agents arranged in a nondecreasing order 
of their saving propensity. The inset compares cumulative distribution of wealth. Parameters are 
 $L=10^4$ and $N=10^4$.}
\label{fig:Compare}
 \end{figure}

  For comparison, we have performed a numerical simulation in a ensemble of $N=10^4$ systems, 
each having $L=10^4$ agents. The saving propensity is chosen from a uniform 
distribution in the interval $(0,1)$ and is ordered in all the systems  such that 
$\lambda_0\le \lambda_1\le \dots \lambda_{L-1}$.  Initial wealth  is also chosen 
randomly in each system with fixed  $\bar E$. Thus the stating wealth 
of any particular agent is  different in different systems. Average wealth 
of each agent $w_i$  is calculated after $10^3$  Monte Carlo steps and then 
the distribution $P(w)$ 
is evaluated. Figure \ref{fig:Compare} compares $w$ and $P(w)$ with exact results.
 
\begin{figure}[tbp] 
\includegraphics*[width=7cm]{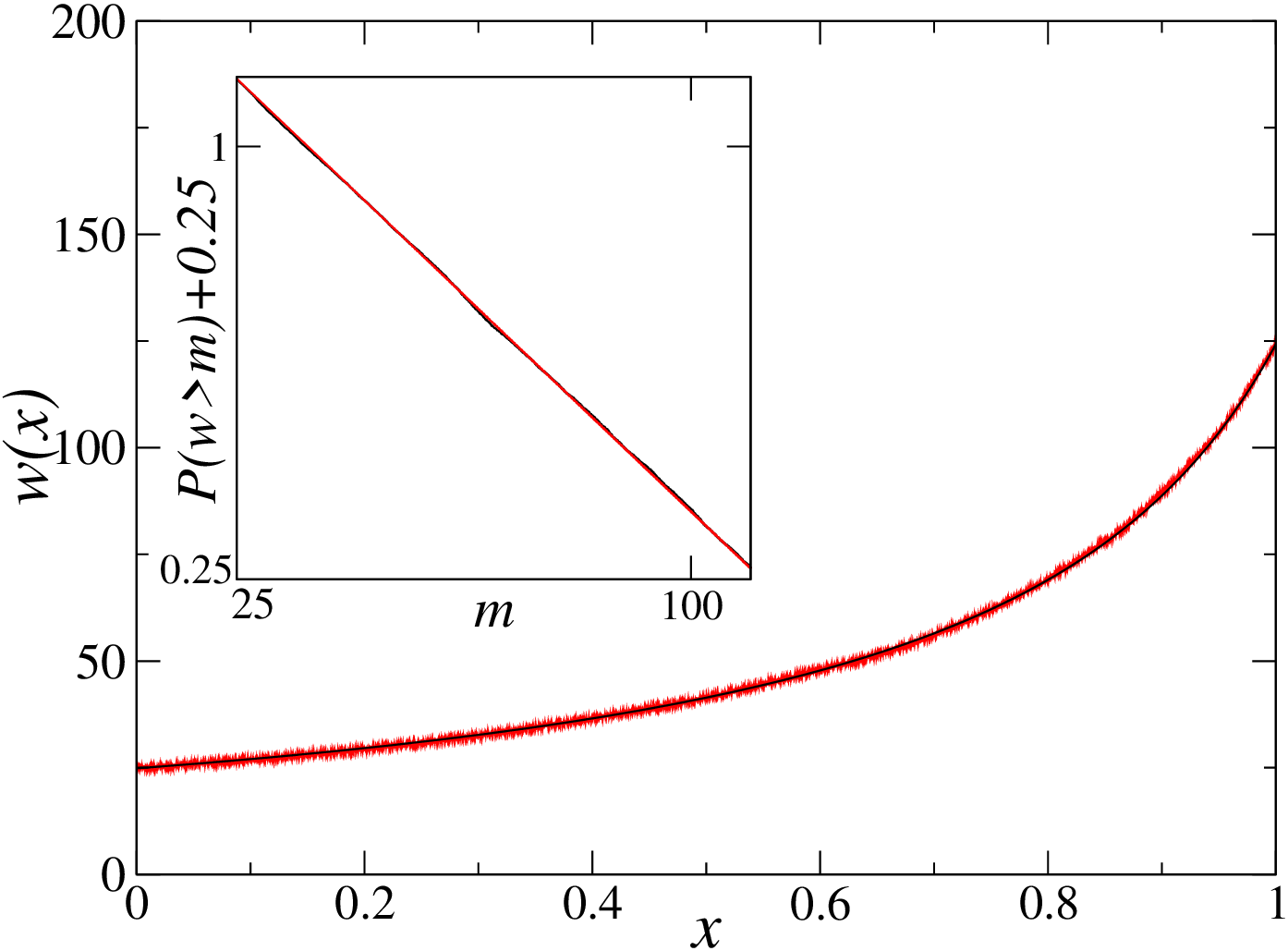}
\caption{$w(x)$ obtained from simulations of a restricted 
saving model ($0<\lambda_i<q$)  is plotted together with the exact result 
(\ref{eq:wq}). Inset compares the cumulative distribution of wealth.   
Here $q=0.8$, $L=10^4=N$, and $\bar E= 50$.}
\label{fig:q}
 \end{figure}

   The results discussed in Eq. (\ref{eq:wq}) are valid for any $q$, 
where one expects $P(w)= C/w^2$. However, earlier numerical studies 
\cite{Bikas03} have reported a power-law distribution only for large $q$. 
The discrepancy can be explained as follows.  The minimum wealth of the system 
is $w(0)= qC$ and the maximum is $w(1)= qC/(1-q^2)$. 
The width of the interval  
$q^3C/(1-q^2)$ is quite small for small $q$  unless $C$ is large.  Since $C$ is 
proportional to $\bar E$ one must 
take large average wealth  to see such a power law. Beacause of the finite lower limit in 
$w$, the cumulative distribution $P(w>m)= 1-q^{-1} +C/n$ shows a power law up to
a constant. We have plotted  $P(w>m)+q^{-1} -1$  vs $m$ in logarithmic scale 
in Fig. \ref{fig:q} which perfectly fits to  $C/m$ (see figure caption for details).

       It is interesting to ask what happens when two different economies interact? 
For example  take a system  of $L$ agents, of which $q$ and $1-q$  fractions  belong to two 
different organizations  $\sigma=\pm$. Their saving propensity distributed 
according to $g_\sigma(\lambda)$, respectively, with $0<\lambda<1$. During  trading 
agent $i$ of type $\sigma$ interacts with agent $j$  (type $\sigma^\prime$) and shares $\epsilon_{ij}(\sigma,\sigma^\prime)$ and  $1- \epsilon_{ij}(\sigma,\sigma^\prime)$ fraction of 
"available wealth", respectively. 
Let us assume that $\langle \epsilon_{ij}(\sigma,\sigma^\prime)\rangle = r_{\sigma \sigma^\prime}$ (obviously $r_{+-}=r_{-+}$).
The distribution  of the grand system is now $ g(\lambda) = q g_+ (\lambda) + p  g_-(\lambda)$ 
and the effective sharing is $r= q^2 r_{++}  + (1-q)^2 r_{--} + 2q(1-q) r_{+-}.$  
No matter  what the value of $r$ is,  Eq. (\ref{eq:wx}) is still valid and thus the distribution of wealth is given by Eq. (\ref{eq:Pw}).  So the power law $P(w) \sim w^{-2}$ and even  Eq. (\ref{eq:wx}) 
are quite robust.

Equation (\ref{eq:wx}) is the central result of this paper,
which states  that the
{\it wealth of an agent having saving propensity $\lambda$ is inversely proportional 
to $1-\lambda$}, irrespective of  what the distribution $g(\lambda)$  and $h(\epsilon)$ are.
It clearly  indicates that on the average each agent, independent of how rich or poor he is, 
invests a constant wealth $C$  (which is of course $(1-\lambda$) fraction of his 
individual wealth) for trading \cite{lessC}.
And then with equal probability he receives $r$ or $1-r$ fraction of the available 
wealth  $2C$. Thus, on the average, the individual wealth is preserved in the 
steady state.

\begin{figure}[tbp] 
 \includegraphics*[width=7. cm]{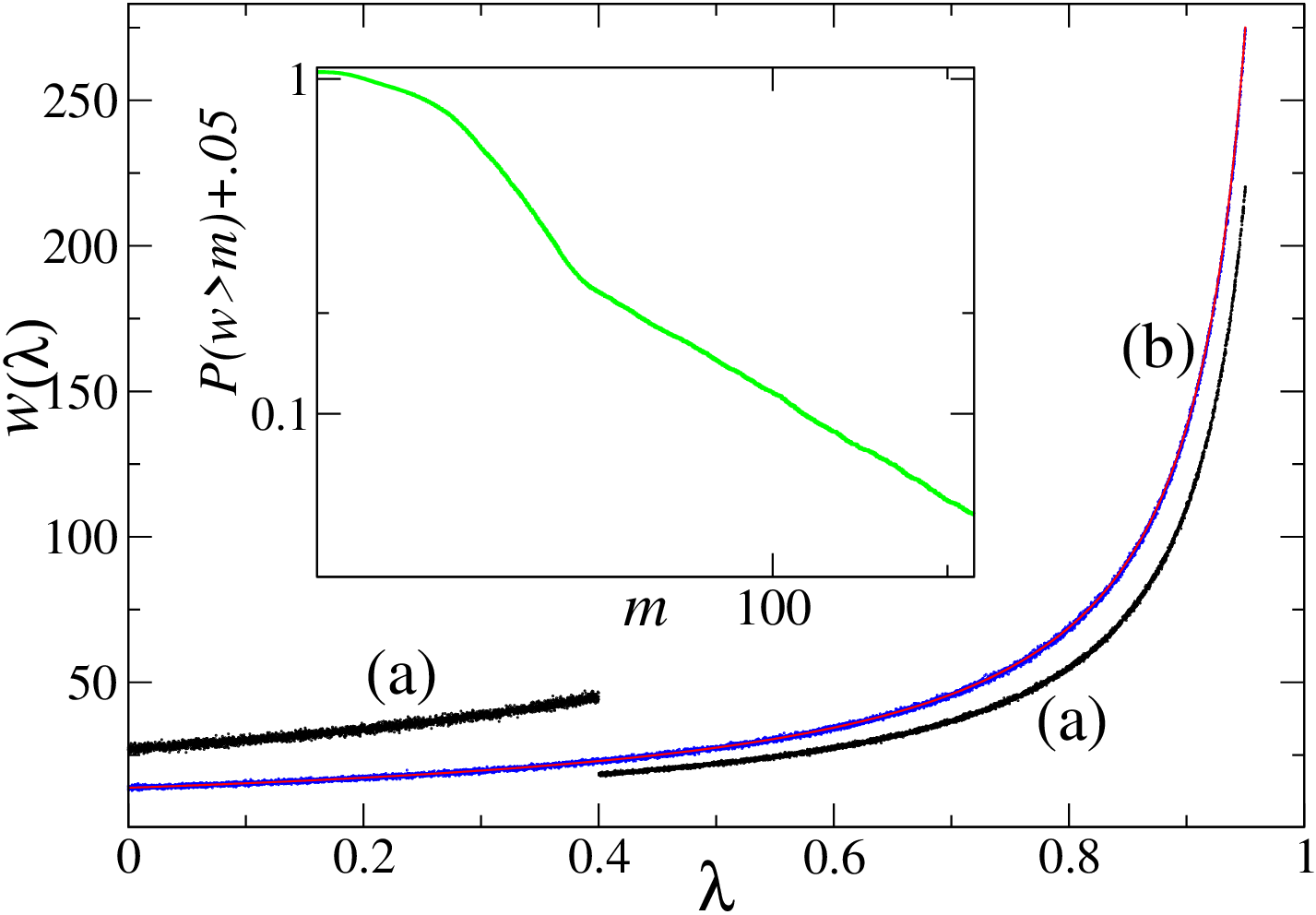}
\caption{(a) $w(\lambda)$ for the \textit{poor-rich} system, where poor savers 
own $2/3$ of the total wealth. Inset shows corresponding cumulative distribution of wealth.
(b) $w(\lambda)$ for the \textit{poor-intermediate-rich} model compared with 
CCM model \cite{Mann04}. Here $z=0.4$, $b=0.6$, $q=0.95$ and,  $L=10^4=N$.}
\label{fig:PoorRich}
 \end{figure}

One can instead write Eq. (\ref{eq:wx}) as
\begin{equation}
w(\lambda) = \frac{C}{1-\lambda} ~~\textrm{with}~~ C^{-1} = \bar E^{-1} \int \frac{g(\lambda)d\lambda}{1-\lambda}
\label{eq:wlambda}
\end{equation}
thus, {\it better saving means better wealth}. To verify the robustness of Eq. 
(\ref{eq:wlambda}) let us divide the system of $L$ agents with their savings distributed 
uniformly in $(0,q)$ into two groups :  (a) the {\it poor} savers who have $\lambda_i<z$ 
and  (b) the {\it rich} savers  who have $z\le\lambda<q$. 
If the poor savers  interact only with the poor  and  the rich savers  interact 
only with the rich, clearly the system breaks up into two independent subsystems 
of size $zL$ and $(q-z)L$, respectively. Correspondingly wealth for the poor and the rich 
are $w_p(\lambda) = C_p/(1-\lambda)$ and  $w_r(\lambda) = C_r/(1-\lambda)$, 
where  $C_p$  and $C_r$ are to be determined independently from  the initial 
average of wealth in each system.  The total distribution  of wealth
is $P(w) = zP_p(w) + (q-z) P_r(w)$, where the distribution for the poor and the rich are 
$P_{r,p}(w)= C_{r,p}/w^2$, nonzero in the interval $w_p(0)<w<w_p(z)$, and $w_r(z)<w<w_p(1)$, 
respectively. Depending on the choice of parameters one can get different 
intervals  where $P_p$, $P_r$ , both or none of them contribute to 
$P(w)$.  For a certain choice, as described in the inset of 
Fig. \ref{fig:PoorRich}, it is possible to obtain a cumulative  wealth distribution which resembles 
the one observed in reality. Correspondingly a  poor-rich break up at $\lambda=0.4$ is shown in 
Fig. \ref{fig:PoorRich}(a). This discontinuity disappears once  some intermediate savers (whose 
saving propensities are limited to $z<\lambda_i<b$)  are introduced  who can interact with both 
the poor and the rich savers. Surprisingly  $w(\lambda)$ and $P(w)$ in this system 
are identical to that of the original CCM model where  every agent interacts with every 
other agent. Fig. \ref{fig:PoorRich}(b) compares $w(x)$ obtained  from  a numerical 
simulation for  the poor-intermediate-rich system with the exact results. 
This suggests that 
that  each agent in the system stands by their own. Irrespective  of their distribution of saving 
propensity and interaction with other agents,  any tagged agent 
who has saving propensity $\lambda$ acquires  wealth which is 
inversely proportional to $(1-\lambda)$. 
Equation (\ref{eq:wlambda}) is  thus quite robust. 

   So far we have discussed  the distribution of  wealth in a ensemble of infinitely many 
identical systems. Instead, if we look at any given system which is a member of the ensemble, 
wealth of any particular agent (saving propensity $\lambda$)  would show fluctuations about 
the average $w= C/(1-\lambda)$.  
Such fluctuations  have been studied in \cite{MannaBook}. These numerical 
studies indicate that the distribution of fluctuations $P(E)$ for any tagged agent (saving propensity $\lambda$) is a Gamma-like distribution which is  asymmetric about the mean $w=C/(1-\lambda)$ and is peaked  at $E_c< w$.  It is also known that the distribution becomes symmetric with  $E_c$ approaching $w$ when  $\lambda\to 1$.  
Since usually agents are crowded around the peak of the distribution, in any particular 
system in the ensemble $P(E)$ is not very different from $P(E_c)$. 
Thus for the rich (who have $\lambda\approx 1 $), the distribution of wealth $P(E)$ is the 
same as $P(w)$  (as $E_c\approx w$), which explains the Pareto law  for the rich  in every system in a 
ensemble \cite{Patri05}. Whereas  deviation of $P(E)$ from a power law is expected for the poor. An 
exact study of fluctuations could reveal the discrepancy.  

What happens in a system of identical agents  (each having same saving 
propensity $\lambda$)?  Note that we need not consider identical copies of the systems now.
The system itself is an ensemble of identical agents. Clearly the average wealth is 
$w= C/(1-\lambda)= \bar E$, and thus the probability distribution  $P(w) = \delta(w-\bar E)$.
Agents in the system differ by their fluctuations and thus the distribution of wealth at any 
given instant  would only count the fluctuations about the average $\bar E$, which is 
not different from the distribution of fluctuations of a tagged agent in CCM model. 
Extensive numerical simulations \cite{Bikas03} have shown a Gamma-like distribution in this case.  
Further analytic study might  shed light on the  exact form of distribution. 
% 
% It is important to notice that the distribution of average wealth $P(w)$  not necessarily 
% dictate  the distribution of fluctuation. For example 
% in a system of identical agents $P(w)$ is a $\delta$-distribution for any $\lambda$ where as 
% $P(E)$  is wildly different for $\lambda=0$ and $\lambda\ne 0$  case. It is  a Gibbs distribution 
% for the former case and a Gamma-like distribution for the later.

In conclusion, we have provided an exact solution to the ideal-gas-like markets using Gibbs 
ensemble approach.  We point out that in a system of nonidentical agents, it is appropriate 
to consider infinitely many identical copies of the systems differing  by  their  initial 
conditions. Such ensembles represent the evolution of several identical systems under the 
same macroscopic conditions and, thus, physical observables  make sense only as an 
ensemble-averaged quantity. One single  system, instead,   would  encounter fluctuations 
which are sometimes incalculably complex.  A system of agents having  the same saving propensity 
is such a case, where average quantities like wealth are identical for every agent and 
fluctuations are the only things that count. 
  The central result revealed from our exact solution is that in ideal-gas-like markets, 
irrespective  of the details of the interaction and distribution of saving propensity,  
an agent having saving propensity $\lambda$ would, on the average, acquire wealth  which is 
inversely proportional to  $1-\lambda$, $i.e.$,  better savers are richer. 
Thus, every agent, poor ( small $\lambda$) or rich (large $\lambda$), on the average 
invests  the same amount for trading, contrary to the real economy where rich agents
usually invest more. To alter such scenario  one might modify this minimal model so that 
the  investment is non-linear in $w$ [note that current investment $(1-\lambda)w$  is linear].

  I would like to acknowledge exciting discussions with B. K. Chakrabarti, 
K. Sengupta, R. Stinchcombe, A. Das, and A. Chatterjee.


\begin{thebibliography}{99}
%\bibitem[*]E E-mail: pk.mohanty@saha.ac.in
\bibitem{Pareto} V. Pareto, Cours d'economie Politique ( F. Rouge, lausanne, 1897).
\bibitem{Real} A. A. Dragulescu and V. M. Yakovenko, Physica {\bf A 299}, 213 (2001);
W. Souma, Fractals 9, 463 (2001); T. Di Matteo, T. Aste and  S. T. Hyde, cond-mat/0310544
and in "{\it The Physics of Complex Systems (New Advances and Perspectives)}", 
Eds. F. Mallamace and H. E. Stanley, (IOS Press, Amsterdam 2004); 
S. Sinha, Physica {\bf A} 555 (2005); F. Clementi and  M. Gallegati, 
Physica {\bf A 350}, 427(2005).
\bibitem{Yako00} A. A. Dragulescu and V. M. Yakovenko,  Eur. Phys. J. {\bf B 17}, 723 (2000).
\bibitem{Bikas03} A. Chakraborti and B. K. Chakrabarti, Eur. Phys. J. {\bf B 17}, 167 (2000).
 
\bibitem{Mann04} A. Chatterjee, B. K. Chakrabarti and S.S. Manna, Physica Scripta T106, 36 (2003);
 \textit{ibid} Physica {\bf A 335}, 155(2004).
\bibitem{Patri04} M. Patriarca, A. Chakraborti and  K. Kaski, \pre {\bf 70}, 016104 (2004).
\bibitem{Rich04}  P. Repetowicz, S. Hutzler and P. Richmond, Physica {\bf A 356}, 641(2005).
\bibitem{Book} {\it Econophysics  of Wealth Distribution}  Eds. 
A. Chatterjee, S. Yarlagadda and  B. K. Chakrabarti, Springer Verlag, Milano 2005.
\bibitem{Das03}  A. Das and S. Yarlagadda, Phys. Scripta \textbf{T106}, 39 (2003); 
A. Chatterjee, B. K. Chakrabarti, R. B. Stinchcombe, Phys. Rev. \textbf{E 72 }, 026126 (2005).
\bibitem{lessC} Note that no one posses lesser wealth than $C$.
\bibitem{MannaBook} K. Bhattacharya, G. Mukherjee and S. S. Manna in  \cite{Book};
Y. Fujiwara {\it et. al.}, Physica {\bf A 321}, 598(2004). 
\bibitem{Patri05} A similar arguement has been used by 
M. Patriarca, A. Chakraborti, K. Kaski and G. Germano in \cite{Book}. 
\end{thebibliography}
\end{document}